\documentclass[lettersize,journal]{IEEEtran}
\usepackage{amsmath,amsfonts,amssymb}
\newtheorem{theorem}{Theorem}
\newtheorem{proof}{Proof}
\usepackage{graphicx}
\usepackage{subfigure}
\usepackage[colorlinks]{hyperref}
\usepackage{multirow,multicol}
\usepackage{makecell}

\begin{document}
\title{\LARGE FAS-Assisted Self-Interference Mitigation \\ for In-Band Full-Duplex Communications}
\author{Hanjiang Hong,~\IEEEmembership{Member,~IEEE}, 
            Kai-Kit Wong,~\IEEEmembership{Fellow,~IEEE},
            Jian Sang,~\IEEEmembership{Member,~IEEE}, \\
            Yiyan Wu,~\IEEEmembership{Life Fellow,~IEEE},
            and Hyundong Shin,~\IEEEmembership{Fellow,~IEEE}
\vspace{-10mm}

\thanks{The work of K. K. Wong is supported by the Engineering and Physical Sciences Research Council (EPSRC) under Grant EP/W026813/1. The work of H. Shin is supported by the National Research Foundation of Korea (NRF) grant funded by the Korean government (MSIT) (RS-2025-00556064 and RS-2025-25442355), and by the Ministry of Science and ICT (MSIT), Korea, under the ITRC (Information Technology Research Center) support program (IITP-2025-RS-2021-II212046), supervised by the IITP (Institute for Information \& Communications Technology Planning \& Evaluation).}
\thanks{H. Hong, K. K. Wong, and J. Sang are with the Department of Electronic and Electrical Engineering, University College London, London, United Kingdom. K. K. Wong is also affiliated with the Department of Electronic Engineering, Kyung Hee University, Yongin-si, Gyeonggi-do 17104, Korea. (e-mail: $\rm \{hanjiang.hong, kai\text{-}kit.wong, jian.sang\}@ucl.ac.uk$).}
\thanks{Y. Wu is with the Department of Electrical and Computer Engineering, Western University, London, ON N6A 3K7, Canada (e-mail: $\rm yiyan.wu@ieee.org$).}
\thanks{H. Shin is with the Department of Electronics and Information Convergence Engineering, Kyung Hee University, Yongin-si, Gyeonggi-do 17104, Republic of Korea (e-mail: $\rm hshin@khu.ac.kr$).}
\thanks{Corresponding author: Kai-Kit Wong.}
}
\maketitle
\begin{abstract}
In-band full-duplex (IBFD) communication relies on effective self-interference cancellation (SIC) to overcome the strong loopback self-interference (SI). While fluid antenna system (FAS)-assisted SIC offers superior SI suppression by dynamically selecting the FAS port to minimize the residual SI power, it incurs high pilot overhead to acquire channel state information (CSI) for both forward and loopback channels. {\color{blue}This raises the question of whether forward channel CSI is necessary for effective SI mitigation. To address this challenge, we propose a pilot-efficient FAS-assisted SI mitigation (FAS-SIM) approach that selects the FAS port solely based on the loopback SI channel.} Theoretical performance analysis of the forward signal is provided. Numerical results then reveal that FAS-SIM achieves SI suppression and a forward-link rate comparable to the conventional method, {\color{blue}while reducing port-selection training overhead by $50\%$,} offering a low-overhead solution for IBFD communications. 
\end{abstract}
\vspace{-1mm}

\vspace{-3mm}
\section{Introduction}
\IEEEPARstart{T}{he evolution} of six-generation (6G) networks demands high spectral efficiency, low latency, and massive throughput to support emerging applications \cite{ATIS,wang2023ontheroad}. In-band full-duplex (IBFD) has emerged as a promising physical-layer technology because it enables simultaneous transmission and reception on the same frequency band, potentially doubling link capacity compared to half-duplex systems \cite{Sabharwal-2010,smida2023FD6G}. A key challenge in IBFD is the severe loopback self-interference (SI). The high-power loopback signal (LBS) from the co-located transmitter can overwhelm the desired forward signal (FWS) from a remote source. Conventional SI cancellation (SIC)  often employs propagation domain isolation, analog cancellation, and digital domain processing \cite{yang2026afreq,le2022atwo,Hong2023frequency} \cite{dong2024augmentation}. 

Fluid antenna systems (FAS) provide an additional spatial degree of freedom through dynamically reconfigurable antenna positions or configurations \cite{wong2021FAS,New2024aTutorial,hong2026acontemp}. The concept was first introduced in 2020 \cite{wong2020FAS}, and has been widely investigated in various channel models \cite{Khammassi2023,hong2025fluid,zhang2026finite}. Recent attempts have also tackled its channel estimation problem \cite{xu2023channel,zhang2026joint}. In an IBFD receiver, the flexibility of FAS can be exploited to select a port experiencing a deep SI fade \cite{tang2026full,sko2023full,hong2026FASSIC}. {\color{blue}Although the co-located transmitter (Tx) and receiver (Rx) geometry is typically fixed, the effective LBS channel may vary with the dynamics of the surrounding scattering environment, motivating dynamic FAS port selection over a fixed-position antenna (FPA) configuration. Under quasi-static conditions, the selected port can be retained to reduce training overhead.} Considering the existing SIC methods, \cite{sko2023full} selects the port with the strongest FWS. The FAS-assisted SIC framework in \cite{hong2026FASSIC} considers both the FWS and LBS channels, achieving substantial SI attenuation. However, \cite{hong2026FASSIC} requires full channel state information (CSI) across all FAS ports for both LBS and FWS channels, in twice the pilot overhead that \cite{sko2023full} requires. In rapidly changing propagation environments or high-density port configurations, transmitting dedicated pilots consumes a prohibitive proportion of the frame structure, severely eroding the gains achieved by IBFD. {\color{blue}Importantly, whether FWS CSI is essential for SI mitigation (SIM) remains unclear.} 

To tackle this challenge, this letter proposes a pilot-efficient FAS-assisted SIM (FAS-SIM) scheme that select the FAS port based merely on the LBS channels, reducing the training pilot overhead compared to \cite{hong2026FASSIC}. Under rich-scattering conditions, a closed-form upper bound and an analytical approximation of the FWS signal-to-interference-plus-noise ratio (SINR) are derived and validated through simulations. Our results show that FAS-SIM achieves nearly identical FWS performance compared to \cite{hong2026FASSIC} {\color{blue}despite requiring only LBS CSI for port selection. Moreover, FAS-SIM reduces the port-selection pilot budget by $50\%$ compared with \cite{hong2026FASSIC}, demonstrating a favorable trade-off between CSI acquisition overhead and SIM performance.} Simulation results under realistic near-field channels \cite{zhang2024NFChan} {\color{blue}further confirm the effectiveness of FAS-SIM, with rich-scattering results serving as an empirical benchmark.}

\begin{figure}
\centering
\vspace{-2mm}
\includegraphics[width = 0.9\linewidth]{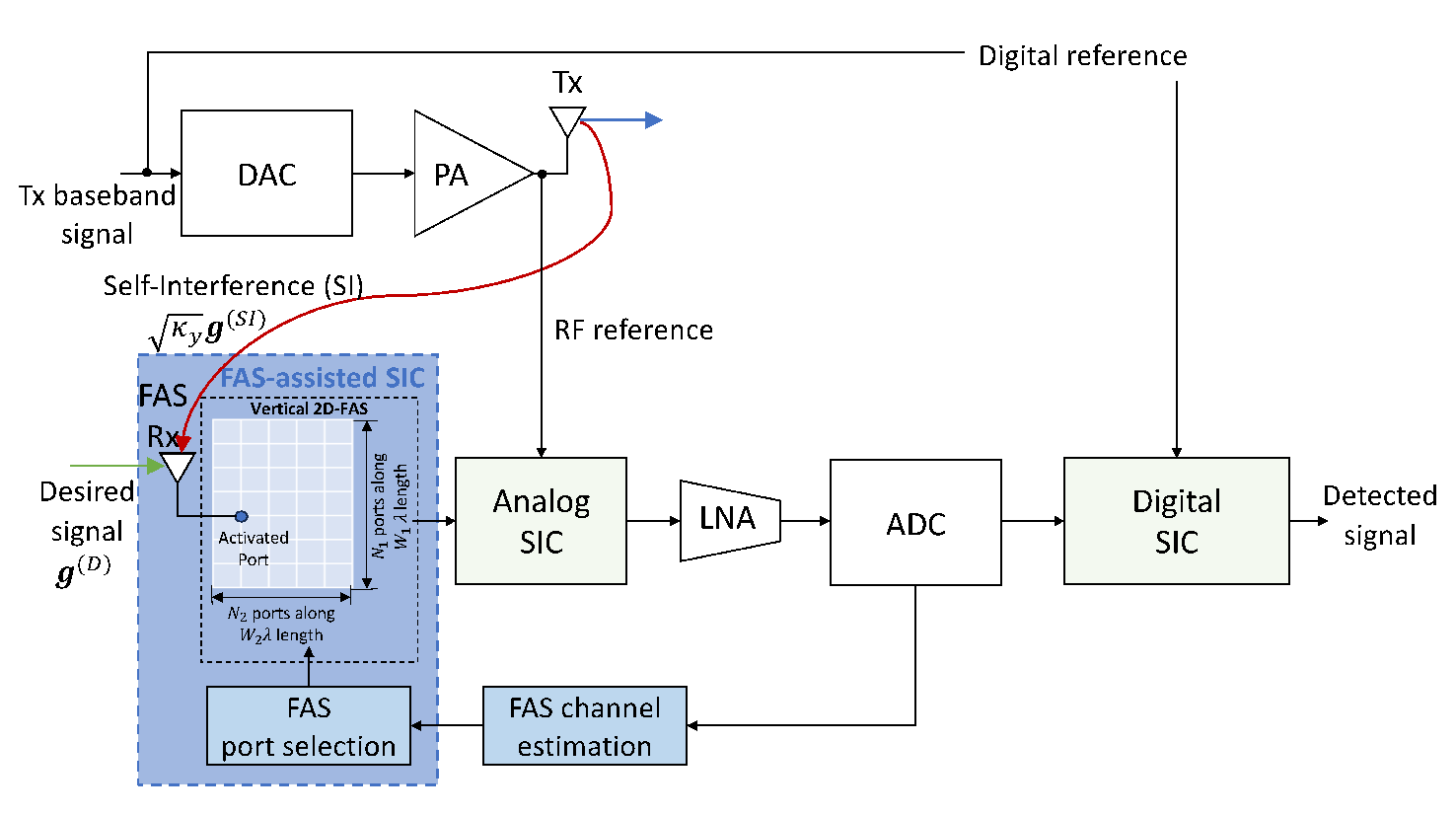}
\vspace{-3mm}
\caption{IBFD system with a FAS-assisted receiver.}\label{Fig:SysModel}
\vspace{-4mm}
\end{figure}

\vspace{-2mm}
\section{FAS-assisted SIC Framework}\label{sec:SystemModel}
As illustrated in Fig.~\ref{Fig:SysModel}, we consider a single-input single-output IBFD communication system with co-located Tx and Rx. The Tx is equipped with an FPA, while the Rx features a two-dimensional FAS (2D-FAS), in which the antenna can be instantaneously switched to one of the $N = (N_1 \times N_2)$ ports uniformly distributed across an area of $W = W_1 \lambda \times W_2 \lambda$, where $\lambda$ denotes the carrier wavelength. The Rx simultaneously receives the desired FWS from the remote Tx and the LBS from the co-located Tx antenna. SIC methods aim to suppress SI in the received signal, thereby enabling processing of the interference-free desired FWS.

{\color{blue}SI is typically mitigated through a combination of analog and digital SIC. The FAS-assisted SIC methods operate in the propagation domain by selecting a port with favorable conditions to receive. This reduces the SI power for subsequent analog and digital SIC stages, relaxing the receiver dynamic-range requirements and the cancellation depth. It complements conventional SIC techniques rather than serves to replace them.} Hypothetically, if all the FAS ports are activated, the received signals with time-index omitted is given by
\begin{equation}\label{Eq:sigmod}
\boldsymbol{y} = \boldsymbol{g}^{({\rm D})} s_{\rm D} +\sqrt{\kappa_y} \boldsymbol{g}^{({\rm SI})} s_{\rm SI}+\boldsymbol{\eta}, 
\end{equation}
in which $s_{\rm D}$ represents the desired FWS, $\boldsymbol{g}^{({\rm D})}$ is the forward channel vector, $s_{\rm SI}$ denotes the LBS, $\boldsymbol{g}^{({\rm SI})}$ represents the LBS channel vector, {\color{blue}$\kappa_y$} is the average power ratio between $\boldsymbol{g}^{({\rm SI})}$ and $\boldsymbol{g}^{({\rm D})}$, and $\boldsymbol{\eta}\sim \mathcal{CN}(0,\sigma_\eta^2\mathbf{I})$ is the additive white Gaussian noise. {\color{blue}For analytical tractability, nonlinear hardware distortions are not explicitly considered in the SI.}

The FAS channels, $\boldsymbol{g}^{({\rm Z})} = [g^{({\rm Z})}_1,\dots,g^{({\rm Z})}_{N}]^T,~{\rm Z}\in \{{\rm D},{\rm SI}\}$, exhibit strong correlation, i.e., $\mathbb{E}\left[ \boldsymbol{g}^{({\rm Z})} (\boldsymbol{g}^{({\rm Z})})^ \dag \right] = \sigma_g^2 \mathbf{\Sigma}$. The $(n,m)$-th element in $\mathbf{\Sigma}$ is characterized as \cite{Khammassi2023} 
\begin{equation}\label{Eq:corr}
\left[\mathbf{\Sigma}\right]_{n,m} \!\!=\! j_0 \! \left(\! 2\pi \sqrt{\! \left(\frac{n_1 \!-\! m_1}{N_1 \!-\! 1} W_1 \!\right)^2 \!\!+\! \left(\frac{n_2 \!-\! m_2}{N_2 - 1} W_2 \! \right)^2}\right),
\end{equation}
where $(n_1, n_2) \to n$, $(m_1, m_2) \to m$, and $j_0(\cdot)$ is the spherical Bessel function of the first kind. By utilizing the eigenvalue-based model, $\boldsymbol{g}^{({\rm Z})}$ under rich scattering can be written as
\begin{equation}\label{Eq:chan_narr_vec}
\boldsymbol{g}^{({\rm Z})} = \sigma_g \mathbf{U} \mathbf{\Lambda}^{\frac{1}{2}} \boldsymbol{a}^{({\rm Z})},
\end{equation}
in which $\boldsymbol{a}^{({\rm Z})} \sim \mathcal{CN} (0,\mathbf{I})$. The matrices $\mathbf{\Lambda}$ and $\mathbf{U}$ stem from the singular value decomposition of $\boldsymbol{\Sigma}$, given by $\mathbf{\Sigma} = \mathbf{U}\mathbf{\Lambda}\mathbf{U}^\dag$. 

We also consider the near-field channel model for the LBS channel $\boldsymbol{g}^{({\rm SI})}$ due to the Tx-Rx co-location, given by \cite{zhang2024NFChan}
\begin{equation}\label{Eq:NFChann}
\boldsymbol{g}^{({\rm SI})}= \sigma_g \left(\sqrt{\!\!\frac{K}{K\!+1\!}}\boldsymbol{g}_{\rm LoS} + \sqrt{\frac{1}{K\!+1\!}}\boldsymbol{g}_{\rm NLoS}\right),
\end{equation}
where $K$ denotes Rice factor and $N_p$ represents the number of scattered components. Moreover, the LoS component is given by ${g}^{\rm LoS}_{n_1,n_2} = e^{j\alpha_{\rm LoS}} \times \exp(-j2\pi r_{n_1,n_2}^{\rm LoS}/\lambda)$. Here, $\alpha_{\rm LoS}$ denotes the random phase and $r_{n_1,n_2}^{\rm LoS}$ represents the distance between Tx and the Rx's $n$-th FAS port, given by
\begin{equation}
r_{n_1,n_2}^{\rm LoS} = \sqrt{\left(\frac{n_1-1}{N_1-1}W_1\lambda\right)^2 + \left(\frac{n_2-1}{N_2-1}W_2\lambda\right)^2 + d_z^2},
\end{equation}
in which $d_z$ denotes the vertical distance between Tx and Rx. The NLoS component is also represented as $\boldsymbol{g}_{\rm NLoS} = \frac{1}{\sqrt{N_p}} \sum_{p=1}^{N_p} a^{\rm NLoS}_p \exp{(-j2\pi r_{n_1,n_2}^{(p)}/\lambda)}$. Each NLoS path has a random coefficient $a^{\rm NLoS}_p$, and the distance $r^{(p)}_{n_1,n_2}$ is calculated by a virtual source located at $\boldsymbol{s}_p = [s_x^{(p)},s_y^{(p)},s_z^{(p)}]^T$.

FAS-assisted SIC methods utilize the spatial diversity of FAS to activate the optimal FAS port. In \cite{sko2023full}, the port that maximizes the FWS channel gain is identified, i.e.,
\begin{equation}\label{Eq:B1_nstar}
n^*_{\rm B1} = \operatorname*{arg\,min}_{n} |g_n^{(\rm D)}|.
\end{equation}
This approach neglects the use of CSI from LBS to mitigate SI effectively. This oversight results in performance degradation within the IBFD system. Recently, \cite{hong2026FASSIC} proposed identifying the FAS port that minimizes the residual SI power, so that
\begin{equation}
n^*_{\rm B2} = \operatorname*{arg\,min}_{n} {|g_n^{(\rm SI)}|^2}/{|g_n^{(\rm D)}|^2}.
\end{equation}
This approach considers both the FWS and LBS channels and provides better performance. However, it also doubles the training pilots for CSI estimation, as it requires full knowledge of both $\{g_n^{(\rm D)}\}_{\forall n}$ and $\{g_n^{(\rm SI)}\}_{\forall n}$ for port selection.

\vspace{-2mm}
\section{Proposed FAS-SIM}\label{sec:SIM}
\subsection{Proposed Method}
We propose a new FAS-assisted SIM method, namely FAS-SIM. Considering the signal model in \eqref{Eq:sigmod}, we can find that the SI component is dominant due to a typically large $\kappa_y$ resulting from co-located Tx and Rx. Therefore, we propose to focus on the SI component and mitigate it by the proposed FAS-SIM method. In FAS-SIM, FAS at the Rx selects the port that mitigates the LBS channel gain. This port selection approach can be expressed mathematically as
\begin{equation}
n^* = \operatorname*{arg\,min}_{n} |g_n^{(\rm SI)}|.
\end{equation}
As such, {\color{blue} the FWS average capacity can be calculated as
\begin{equation}\label{eq:C}
C = \mathbb{E}\left[\log_2\left(1+\frac{E_s|g_{n^*}^{(\rm D)}|^2}{\kappa_yE_{\rm SI}|g_{n^*}^{(\rm SI)}|^2+\sigma_\eta^2}\right)\right].
\end{equation}
Moreover, the} SINR of the desired FWS is given by 
\begin{align}\label{Eq:DefSINR}
\gamma 
& = \frac{E_s\sigma_g^2}{\kappa_y E_{\rm SI} \mathbb{E}\{|g_{n^*}^{(\rm SI)}|^2\} +\sigma_\eta^2} \overset{(a)}{\approx} \frac{E_s\sigma_g^2}{\kappa_y E_{\rm SI}\mathbb{E}\{|g_{n^*}^{(\rm SI)}|^2\}},
\end{align} 
where $\mathbb{E}\{|s_{\rm D}|^2\} = E_s$, $\mathbb{E}\{|s_{\rm SI}|^2\} = E_{\rm SI}$, and $(a)$ is obtained from the fact that the noise power is considerably less than the interference power due to a large $\kappa_y$.

\vspace{-3mm}
\subsection{Performance Analysis}\label{sec:Perf}
Here, our aim is to analyze the performance of FAS-SIM under rich scattering. To simplify the notation, the following definitions are introduced:
\begin{equation}
X_n \triangleq {|g_{n}^{(\rm SI)}|^2}/{\sigma_g^2}, ~
Y \triangleq \min \{X_1, \dots, X_N\}. 
\end{equation}
The SINR of the desired FWS is thus $\gamma = {E_s}/{(\kappa_yE_{\rm SI}\mathbb{E}\{Y\})}$.

We first provide an upper bound on the FWS SINR of the FAS-SIM. For simplicity in computation, this upper bound does not account for the correlation among the ports and indicates the idealized SIC capability of the proposed method. 
\begin{theorem}\label{th:ubSINR}
The FWS SINR, ${\gamma}$, is upper bounded by
\begin{equation}\label{Eq:ubSINR}
{\gamma}_{ub} = \frac{NE_s}{\kappa_y E_{\rm SI}}.
\end{equation}
\end{theorem}

\begin{proof}
See Appendix \ref{app:ProofTheo1}.
\end{proof}

The upper bound $\gamma_{ub}$ can be viewed as an ideal special case of $\gamma$ that lacks spatial correlation among the ports. In this context, an increase in the number of ports increases the SINR, enhancing SIC capability and strengthening FWS. However, considering FAS with a limited configuration, the ports are usually correlated. Accordingly, with a fixed $N$, the SINR $\gamma$ will approach the upper bound $\gamma_{ub}$ if $W$ is large enough.

Subsequently, we take the correlation into consideration and employ the first-stage approximation delineated in \cite{Khammassi2023} to approximate the LBS channel. In this approximation, only $M \ll N$ terms with the largest eigenvalues are considered. The approximated channel, denoted as $\hat{g}_n^{({\rm SI})}$, is defined as
\begin{equation}\label{Eq:g_fsapprox}
\hat{g}_n^{({\rm SI})} \!\!=\! \sigma_g \!\! \left(\sum_{m = 1}^{M} \!\! \sqrt{\lambda_m}\mu_{n,m} a_m \!+\! \sqrt{1\!-\!\!\sum_{m = 1}^{M} \lambda_m \mu^2_{n,m}} b_n\right)\!\!,
\end{equation}
where {\color{blue}$\lambda_m = [\mathbf{\Lambda}]_{m,m}$, $\mu_{n,m}=[\mathbf{U}]_{n,m}$}, the superscript of $a_m^{(\rm SI)}$ is omitted for simplicity and the variable $b_n \sim \mathcal{CN} (0,1)$ is introduced to ensure a constant normalized variance of the channel $\hat{g}^{({\rm SI})}_n$. Thus, the SINR can be approximated as 
\begin{equation}\label{Eq:aveSINR_fsapprox}
\gamma \approx \frac{E_s}{\kappa_y E_{\rm SI} \mathbb{E}\{\hat{Y}\}},
\end{equation}
where the random variables {\color{blue}$\hat{X}_n$} and $\hat{Y}$ are defined as
\begin{equation}
{\color{blue}\hat{X}_n} = {|\hat{g}^{(\rm SI)}_n|^2}/{\sigma^2_g},~\hat{Y} = \min\left\{\hat{X}_1,\dots,\hat{X}_N\right\}.
\end{equation}

\begin{theorem}\label{th:approx}
With the approximation $\hat{g}_n^{({\rm SI})}$ in \eqref{Eq:g_fsapprox}, the expectation of $\hat{Y}$ is given by \eqref{Eq:ExpYApprox} {at the top of the next page}, respectively, where $Q_1(\cdot,\cdot)$ denotes the Marcum Q-function of order 1, and $\alpha_n$ and $\beta_n$ are given by
\begin{equation}\label{Eq:alphabeta}
    \left\{
        \begin{array}{l}
            \alpha_n = \left|\sum\limits_{m=1}^{M} \sqrt{\lambda_m}\mu_{n,m} a_m\right|^2, \\
            \beta_n = \frac{1}{2} \left(1-\sum\limits_{m=1}^{M} \lambda_m \mu_{n,m}^2\right).
        \end{array}
    \right.
\end{equation}
\end{theorem}
\begin{figure*}[tb]
    \begin{equation}\label{Eq:ExpYApprox}
            \mathbb{E} \{\hat{Y}\} =  \frac{1}{\pi^{M}} \int_0^{+\infty} \iint_\mathbb{C}  \cdots \iint_\mathbb{C}  \exp{\left( - \sum_{m=1}^{M} \lvert a_m\rvert ^2\right)} \times \prod_{n=1}^{N} {Q_1 \left(\sqrt{\frac{\alpha_n}{\beta_n}},\sqrt{\frac{y}{\beta_n}}\right)} da_1 \cdots da_{M} {\color{blue}dy}
    \end{equation}
    \hrulefill
\end{figure*}

\begin{proof}
See Appendix \ref{app:ProofTheo2}.
\end{proof}

By considering the $M$ largest eigenvalues, the LBS channel model has been significantly simplified. {\color{blue}An analytical approximation} for the FWS SINR is provided for corroborating the simulation results. Specifically, based on \eqref{Eq:aveSINR_fsapprox} and Theorem~\ref{th:approx}, we can obtain an approximation of FWS SINR $\gamma$ for the IBFD system, and evaluate the effectiveness of the SIC capability enabled by the proposed FAS-SIM method.

\vspace{-2mm}
\section{Simulation Results}\label{sec:Sim}
In this section, we present simulation results to evaluate the performance of the proposed FAS-SIM method in the IBFD system. For simplicity, we assume that the normalized signal power, normalized constant nonlinear SI power, and normalized channel gains are set to $E_s = 1$, $E_{\rm SI} = 1$, and $\sigma_g^2 = 1$. The power ratio is set to $\kappa_y = 30~{\rm dB}$. We compare our proposed FAS-SIM method with both the FPA-based system and two FAS-assisted benchmarks: \emph{Benchmark 1} refers to the FAS-assisted method that maximizes the FAS in \cite{sko2023full}, and \emph{Benchmark 2} represents the FAS-assisted SIC method that minimizes the residual SI in \cite{hong2026FASSIC}. In the FPA system, the fixed antenna cannot mitigate loopback SI, thereby also serving as a benchmark for performance before cancellation.

\begin{figure}[t]
\centering
\vspace{-3mm}
\includegraphics[width = 0.75\linewidth]{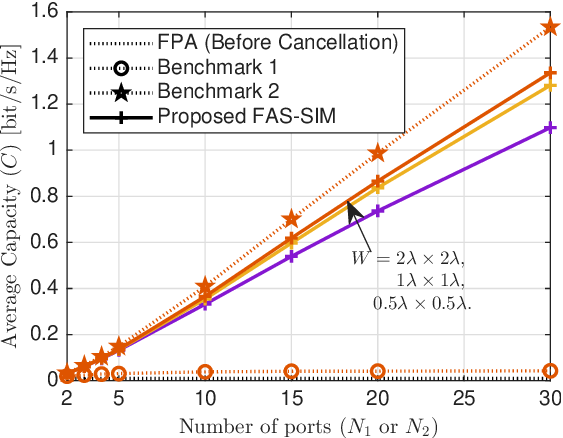}
\vspace{-3mm}
\caption{Average capacity in rich scattering channels.}\label{Fig:RatevsNandW}
\vspace{-3mm}
\end{figure}

Fig.~\ref{Fig:RatevsNandW} shows the desired FWS average capacity. {\color{blue}The capacity is calculated as in \eqref{eq:C}}. For \emph{Benchmark 1 \& 2}, the results are presented for a large FAS with $W = 2\lambda \times 2\lambda$. The capacity of the FPA-based system is close to $0~{\rm bit/s/Hz}$, signifying that the desired FWS is entirely overwhelmed by the transmitted LBS before SIC. The capacity of \emph{Benchmark 1} exceeds that of FPA, but remains close to $0~{\rm bit/s/Hz}$. In contrast, \emph{Benchmark 2} approach yields a satisfactory average capacity for FWS, while requiring twice the training pilots for both $\{g^{({\rm D})}_n\}$ and $\{g^{({\rm SI})}_n\}$. The proposed FAS-SIM method provides a good trade-off. Its performance approaches \emph{Benchmark 2} while having the same training pilot overhead as in \emph{Benchmark 1}. The FWS capacity of FAS-SIM increases with $N$, indicating that increasing $N$ within a constrained FAS size enhances FWS. Moreover, capacity also increases with $W$ but reaches a saturation point. For instance, the red solid curve representing $W = 2\lambda \times 2\lambda$ exhibits marginal performance improvement compared to the yellow curve with $W = \lambda\times \lambda$. 

\begin{figure}[t]
\centering
\vspace{-3mm}
\includegraphics[width = 0.73\linewidth]{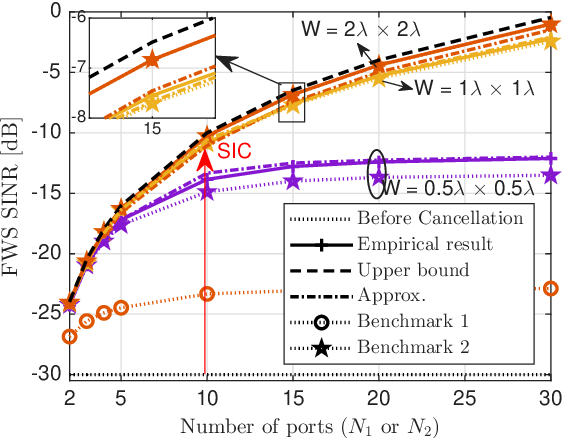}
\vspace{-3mm}
\caption{FWS SINR for different FAS configurations with various sizes and resolutions in rich scattering channels.}\label{Fig:SINRvsNandW} \vspace{-3mm}
\end{figure}

Fig.~\ref{Fig:SINRvsNandW} illustrates the analysis in Section \ref{sec:Perf}. The FWS SINR, $\gamma$, is defined in \eqref{Eq:DefSINR}, with its upper bound calculated by \eqref{Eq:ubSINR} and its approximation obtained by \eqref{Eq:aveSINR_fsapprox} and Theorem \ref{th:approx}. The expectation in Theorem \ref{th:approx} is calculated based on the numerical integral of $\hat{\boldsymbol{a}}$ with $M = 10$. The empirical results approach the upper bound as $W$ increases. For $W = 0.5\lambda \times 0.5\lambda$ or $\lambda \times \lambda$, the approximation is deemed sufficiently accurate. However, a small gap persists between the approximation and the empirical results when $W = 2\lambda \times 2\lambda$, indicating that a higher approximation level of $M$ is necessary to achieve the desired accuracy for a large $W$.

The SINR improvement of the FAS-assisted methods compared to the SINR before cancellation ($-30~{\rm dB}$) can be interpreted as the SIC capability. As can be seen,  the proposed FAS-SIM exhibits much superior and comparable SIC capabilities in comparison to \emph{Benchmark 1} and \emph{Benchmark 2}, respectively. With a small size of $W = 0.5\lambda \times 0.5\lambda$, FAS-SIM can achieve a SIC of approximately $15~{\rm dB}$ with a sufficiently large $N$, and it increases to approximately $28~{\rm dB}$ with a larger $W$.
For a fixed $W$, the SINR of FAS-SIM increases with $N$ or $W$ until a saturation point is reached. In essence, the SINR for $W = \lambda \times \lambda$ and $2\lambda \times 2\lambda$ is comparable and closely approximates the upper bound, suggesting that a regular FAS size is sufficient to provide substantial SIC.

\begin{figure}[t]
\centering
\vspace{-3mm}
\includegraphics[width = 0.75\linewidth]{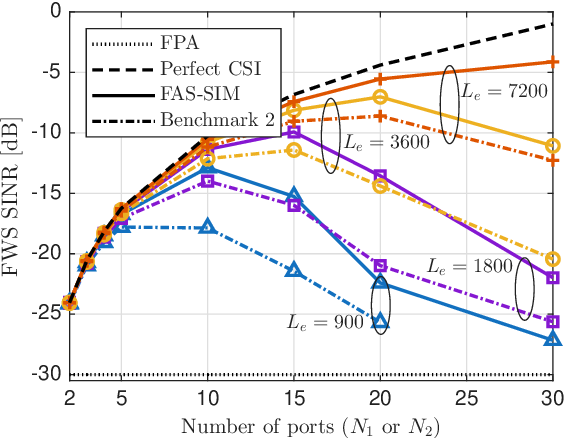}
\vspace{-3mm}
\caption{FWS SINR for different length of channel estimation symbols, $L_e$, with the FAS size of $W = 2\lambda \times 2\lambda$.}\label{Fig:SINRvsN_CE}\vspace{-3mm}
\end{figure}

Fig.~\ref{Fig:SINRvsN_CE} evaluates the impact of channel estimation on performance. The linear minimum-mean-squared-error (LMMSE)-based channel estimation \cite{sko2023full} is considered here. Specifically, $L_e$ pilot-training symbols are employed to estimate $\{{g}^{({\rm SI})}_n\}_{\forall n}${\color{blue}, with $(L_e/N)$ symbols uniformly allocated to each port}. We also present the results of \emph{Benchmark 2} for comparison, in which $L_e/2$ symbols are used to estimate $\{{g}^{({\rm SI})}_n\}_{\forall n}$ and $L_e/2$ symbols are for $\{{g}^{({\rm D})}_n\}_{\forall n}$. As can be observed, FAS-SIM outperforms \emph{Benchmark 2} with the same $L_e$. Moreover, FAS-SIM with $L_e$ training symbols provides similar performance to \emph{Benchmark 2} with $2L_e$ training symbols, because \emph{Benchmark 2} requires twice the training symbols to estimate both $\{{g}^{({\rm SI})}_n\}_{\forall n}$ and $\{{g}^{({\rm D})}_n\}_{\forall n}$. Also, increasing $N$ reduces performance when using a limited $L_e$, as the allocation of pilot-training symbols for each port diminishes, thereby impairing channel estimation quality and jeopardizing the performance. Conversely, by increasing $L_e$, sufficient symbols can be allocated for channel estimation of all FAS ports. As a result, the performance increases, approaching the ideal scenario with perfect CSI. 

\begin{figure}[t]
\centering
\includegraphics[width = 0.75\linewidth]{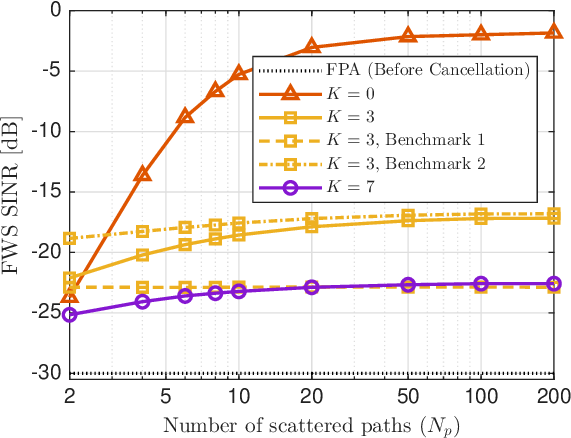}
\vspace{-3mm}
\caption{FWS SINR for different channel condition $(K,N_p)$, in near field channels, with FAS configuration of $N = 30 \times 30$ and $W = 2\lambda \times 2\lambda$.}\label{Fig:SINRvsNp_NFChan}
\vspace{-3mm}
\end{figure}

Fig.~\ref{Fig:SINRvsNp_NFChan} presents the results utilizing the practical near-field channel model for LBS. The vertical distance between the Tx and Rx is set to $\lambda/2$. {\color{blue} The results of \emph{Benchmark 1 \& 2} are presented for comparison.} As can be seen, approximately $28~{\rm dB}$ SIC can be achieved when LoS is absent and multipath is rich, which aligns closely with the results in Fig.~\ref{Fig:SINRvsNandW}. This indicates that the SIC capability in rich scattering channels may serve as {\color{blue}an empirical benchmark} for near-field scenarios. The SINR diminishes with a decrease in $N_p$ and an increase in $K$, illustrating that the SIC capability of FAS-SIM deteriorates in strong LoS environments. {\color{blue}Moreover, FAS-SIM performs comparably to \emph{Benchmark 2} when $N_p$ is large, but performs less effectively when $N_p$ is low, indicating that it is better suited to rich-scattering environments.}

\vspace{-2mm}
\section{Conclusion}\label{sec:conclusion}
This paper proposed FAS-SIM, a pilot-efficient SIM scheme for IBFD communications that selects the FAS port solely based on the loopback SI channel. Analytical expressions for the upper bound and an approximation of the FWS SINR after FAS-SIM were derived under rich-scattering channels. Numerical results under both rich-scattering and realistic near-field channels demonstrated that FAS-SIM achieves comparable SI suppression performance to the existing approach while reducing port-selection training overhead by $50\%$. FAS-SIM thus offers a scalable solution to unlock the full spectral efficiency potential of IBFD in 6G networks.

\vspace{-3mm}
\appendices
\section{Proof of Theorem \ref{th:ubSINR}}\label{app:ProofTheo1}
Applying the unitary matrix $\mathbf{U}$ in \eqref{Eq:chan_narr_vec} to $\boldsymbol{y}$, we have
\begin{equation}
    \boldsymbol{y}' = \mathbf{U}^\dag \boldsymbol{y} = \boldsymbol{h}^{({\rm D})}s_{\rm D} +\sqrt{\kappa_y}\boldsymbol{h}^{({\rm SI})} s_{\rm SI} +\boldsymbol{\eta}',
\end{equation}
where
\begin{equation}\label{Eq:hh}
    \boldsymbol{h}^{({\rm Z})} = \mathbf{U}^\dag \boldsymbol{g}^{({\rm Z})} = \sigma_g \mathbf{\Lambda}^{\frac{1}{2}}\boldsymbol{a}^{({\rm Z})}, ~ {\rm Z} \in \{{\rm D},{\rm SI}\},
\end{equation}
and $\boldsymbol{\eta}' = \mathbf{U}^\dag \boldsymbol{\eta} \sim \mathcal{CN} (\boldsymbol{0}, \sigma_\eta^2 \boldsymbol{I})$. Following this de-correlated transformation, $\boldsymbol{h}^{({\rm Z})} \sim \mathcal{CN} (\boldsymbol{0},\sigma_g^2 \boldsymbol{\Lambda})$, and the entries in $\boldsymbol{h}^{({\rm Z})}$ are independent of each other.

From \eqref{Eq:hh}, we can derive the SINR of FWS in the de-correlated channel as
\begin{equation}
    \gamma' = \frac{E_s\mathbb{E}\{|h_{n^*}^{({\rm D})}|^2\}}{\kappa_y E_{\rm SI}\mathbb{E}\{|h_{n^*}^{({\rm SI})}|^2\}} = \frac{E_s\mathbb{E}\{|a_{n^*}^{({\rm D})}|^2\}}{\kappa_y E_{\rm SI}\mathbb{E}\{|a_{n^*}^{({\rm SI})}|^2\}}.
\end{equation}
For all $1\leq n \leq N$ and ${\rm Z} \in \{{\rm D},{\rm SI}\}$, it can be seen that $a_n^{({\rm Z})}$ are independent of each other, $|a_n^{({\rm Z})}| \sim {\rm Rayleigh}\left(1/\sqrt{2}\right)$, and $|a_n^{({\rm Z})}|^2 \sim {\rm Exponential}\left(1\right)$. 
Evidently, $\gamma'$ can be interpreted as the SINR in an uncorrelated channel represented by
\begin{equation}\label{Eq:y_vector_uc}
    \tilde{\boldsymbol{y}} = \boldsymbol{a}^{({\rm D})}s_{\rm D} +\sqrt{\kappa_y} \boldsymbol{a}^{({\rm SI})}s_{\rm SI} + \tilde{\boldsymbol{\eta}}.
\end{equation}
This uncorrelated channel can additionally be regarded as a special case of that in \eqref{Eq:sigmod}, applicable when ports are sufficiently distanced such that the channel gains across different ports are uncorrelated. Thus, the cumulative density function (CDF) of $Y$ is upper bounded by that of $Y'$, and $\gamma'$ serves as an upper bound for $\gamma$.

Now, denote
\begin{equation}\label{Eq:XXYY}
    X'_n \triangleq |a_n^{(\rm SI)}|^2,~
    Y' \triangleq \min\{X'_1, \dots,X'_N\}, 
\end{equation}
and the FWS SINR of the uncorrelated channel is expressed as ${\gamma}' = {E_s}/{(\kappa_yE_{\rm SI}\mathbb{E}\{Y'\})}$. The CDF of $Y'$ is calculated as
\begin{equation}
F_{Y'}(y) = 1 - \left(1-F_{X_n}(y)\right)^N
\overset{(a)}{=} 1-\exp{(-Ny)}\label{Eq:CDFRR_min_proof}
\end{equation}
where $(a)$ uses the CDF exponential distribution for $X_n$. 
The expectation of $Y'$ is calculated as
\begin{align}
\mathbb{E}\left\{ Y' \right\} & = \int_0^{+\infty} {\left[1-F_{Y'}(y)\right]} dy = \frac{1}{N}.\label{Eq:expRR_min}
\end{align}
By substituting \eqref{Eq:expRR_min} into ${\gamma}'$, we obtain \eqref{Eq:ubSINR}.

\vspace{-2mm}
\section{Proof of Theorem \ref{th:approx}}\label{app:ProofTheo2}
It is established from \eqref{Eq:g_fsapprox} that the approximated channel $\hat{g}^{({\rm SI})}_n$ adheres to a Gaussian distribution. Given a specific $\hat{\boldsymbol{a}} = [a_1,\dots,a_{M}]^T$, we have
\begin{equation}
\hat{g}^{({\rm SI})}_n \sim \mathcal{CN} \left(\sigma_g \sum_{m=1}^{M} \sqrt{\lambda_m} \mu_{n,m} a_m, 2\sigma^2_g\beta_n \right),
\end{equation}
where $\beta_n$ is given in \eqref{Eq:alphabeta}. Hence, $\sqrt{\hat{X}_n} = \lvert \hat{g}^{({\rm Z})}_n \rvert/\sigma_g$ follows Rice distribution, i.e., $\sqrt{\hat{X}_n} \sim {\rm Rice} \left( \sqrt{\alpha_n}, \sqrt{\beta_n} \right)$, where $\alpha_n$ and $\beta_n$ are given in \eqref{Eq:alphabeta}. We now consider the random variable of $\hat{X}_n$ conditioned on $\hat{\boldsymbol{a}}$. This variable follows a non-central Chi-square distribution, with CDF  expressed as 
\begin{equation}
F_{\hat{X}_n | \hat{\boldsymbol{a}}}(x) = 1-Q_1\left(\sqrt{\frac{\alpha_n}{\beta_n}},\sqrt{\frac{x}{\beta_n}}\right).
\end{equation}
The independent variables $b_n, \forall n \in \{1,\dots, N\}$ in \eqref{Eq:g_fsapprox} indicate that $\hat{X}_n$ are independent of each other. Consequently, the CDF of $\hat{Y} = \min\{\hat{X}_1,\dots, \hat{X}_{N}\}$ can be calculated as
\begin{align}
F_{\hat{Y}|\hat{\boldsymbol{a}}}(y) &= 1- \prod_{n=1}^{N}\left[1-F_{\hat{X}_n|\hat{\boldsymbol{a}}}(y)\right]\notag\\ 
&= 1- \prod_{n=1}^{N}{Q_1\left(\sqrt{\frac{\alpha_n}{\beta_n}},\sqrt{\frac{y}{\beta_n}}\right)}. \label{Eq:conPDF_hatY}
\end{align}
The expectation of $\hat{Y}$ conditioned on $\hat{\boldsymbol{a}}$ is expressed as
\begin{align}
\mathbb{E} \{\hat{Y}| \hat{\boldsymbol{a}} \} = & \int_{0}^{+\infty} 1- F_{\hat{Y}|\hat{\boldsymbol{a}}}(y) dy \notag\\ 
= & \int_{0}^{+\infty} \prod_{n=1}^{N} {Q_1\left(\sqrt{\frac{\alpha_n}{\beta_n}},\sqrt{\frac{y}{\beta_n}}\right)} dy.\label{Eq:conExp_hatY}
\end{align}
Thus, the expectation of $\hat{Y}$ can be obtained by the expectation of \eqref{Eq:conExp_hatY} over $\hat{\boldsymbol{a}}$, leading to the result in \eqref{Eq:ExpYApprox}.

\vspace{-2mm}
\bibliographystyle{IEEEtran}

\end{document}